\documentclass[preprint]{aastex}

\def\Msun{{\,M_\odot}}

\begin{document}

\title{Dynamical Friction in dE Globular Cluster Systems}

\author{Jennifer M. Lotz \altaffilmark{1}, Rosemary Telford \altaffilmark{2},
Henry C. Ferguson \altaffilmark{3}}
\author{ Bryan W. Miller \altaffilmark{4}, Massimo Stiavelli \altaffilmark{3}
and Jennifer Mack \altaffilmark{3}}

\altaffiltext{1}{Henry A. Rowland Department of Physics and Astronomy,
Johns Hopkins University, 3400 N. Charles St., Baltimore, MD 21218; 
jlotz@pha.jhu.edu}
\altaffiltext{2}{Department of Physics and Astronomy, University of Wales, Cardiff, Wales, U.K. CF2 3YB; TelfordRE@Cardiff.ac.uk}
\altaffiltext{3}{Space Telescope Science Institute, 3700 San Martin Dr.,
Baltimore, MD 21218;\\ ferguson@stsci.edu, mstiavel@stsci.edu, mack@stsci.edu}
\altaffiltext{4}{Gemini Observatory, Casilla 603, La Serena, Chile;
bmiller@gemini.edu}

\begin{abstract}

The dynamical friction
timescale for globular clusters to sink to the center of a dwarf elliptical
galaxy (dE) is 
significantly less than a Hubble time if the halos have King-model or
isothermal profiles and the globular clusters formed with the same
radial density profile as the underlying stellar population. 
We examine the summed radial distribution of the entire globular cluster 
systems and the bright globular cluster candidates in
51 Virgo and Fornax Cluster dEs for evidence
of dynamical friction processes.  We find that the summed distribution
of the entire globular cluster population closely follows the exponential
profile of the underlying stellar population. However, there is a deficit
of bright clusters within the central regions of dEs (excluding 
the nuclei), perhaps due to the orbital decay of these massive clusters
into the dE cores.  We also predict the magnitude of each dE's nucleus assuming
the nuclei form via dynamical friction. The observed trend of
 decreasing nuclear luminosity with decreasing
dE luminosity is much stronger than predicted if the nuclei formed via simple
dynamical friction processes. We find that the bright dE nuclei
could have been formed from the merger of orbitally decayed massive clusters,
but the faint nuclei are several magnitudes fainter than expected.  These 
faint nuclei are found primarily in $M_V > -14$ dEs which have high globular
cluster specific frequencies and extended globular cluster systems.  
In these galaxies, supernovae-driven winds, high central dark matter densities,
extended dark matter halos, the formation of new star clusters, 
or tidal interactions
may act to prevent dynamical friction from collapsing the entire globular
cluster population into a single bright nucleus.

\end{abstract}


\section{Introduction}
Dwarf elliptical galaxies (dEs) are small, 
low mass ($\sim$ $10^9$ $\Msun$), 
low luminosity ($M_B > -16$), elliptical shaped galaxies
with little or no gas and no ongoing star formation.  
They are the most common type of galaxy in the local universe, outnumbering
giant galaxies and dwarf irregulars in nearby clusters (Binggeli, Sandage \&
Tammann 1985;
Ferguson \& Sandage 1988) and appearing in roughly the same numbers
as dwarf irregulars in low-density environments such as the Local
Group (Grebel 1997).  
In ``bottom-up'' hierarchical merging scenarios
of galaxy formation, dwarf-sized objects are the fundamental building
blocks of larger galaxies. 
If dEs formed via gravitational collapse
onto compact dark matter halos, then they may present a fossil
record of the seeds of giant galaxy formation.  On the other hand, if dEs
form from the tidal debris of galaxy collisions (Mirabel, Dottori, \& Lutz 
1992; Barnes \& Hernquist 1992; 
Hunsberger, Charlton, \& Zaritsky 1996), then they
are but a side-effect of giant galaxy formation. A better understanding
of dE star formation histories and the nature of dE dark matter halos
can help distinguish between these different formation scenarios.

The variation of structure and chemical abundance with dE galaxy mass
is most simply understood as the result of rapid star-formation within 
a dark matter halo, terminated by supernova-driven winds (Dekel \& Silk 1986).
However, the varied star-formation histories of the Local Group dwarf
spheroidal (dSph) and dE galaxies
suggest that the picture is more complicated, at least for these nearby
examples (Grebel 1997).  Constraints on the ages and metallicities of the more
extensive dE populations in nearby clusters are somewhat ambiguous,
although integrated optical and infrared colors and spectral features
(Thuan 1985; Caldwell \& Bothun 1987; James 1994; Bothun \& Mould 1988)
 suggest that dEs are primarily old, metal-poor populations with 
a few exceptions.
Environmental effects, such as tidal interactions, galaxy harassment, 
and ram-pressure stripping, may play a large role in the transformation 
of gas-rich proto-dEs into the gas-poor dEs and regulate their star
formation processes (e.g. Moore, Lake \& Katz 1998).

Brighter than $M_B = -14$, over 50\% of dEs possess compact
stellar nuclei  (Sandage, Binggeli \& Tammann 1985).  The
nuclei for the most part appear as unresolved sources (even at HST
resolution) projected very near the isophotal center of the galaxy. These
could, at least in principle, be the merged conglomeration of globular
clusters that have spiraled into the centers of the galaxies  (Tremaine,
Ostriker, \& Spitzer 1975). However, there are reasons to suspect that
dE nucleus formation may not be so straight-forward.   
Within  the Virgo and Fornax clusters, the spatial distributions of the 
dE,N galaxies and dE galaxies brighter than $M_B = -14$ are 
significantly different.
The dE,N galaxies follow the radial distribution of the early-type (E and S0s)
giant galaxies, while the bright 
dEs (no N) galaxies follow the radial distribution
of the spirals and irregulars (Ferguson \& Sandage 1989).  
It is difficult to see how a purely
internal process for the formation of dE nuclei such as dynamical friction
could mimic such an 
environmental effect.  Also, the dE,N galaxies have a somewhat higher
globular cluster specific frequency $S_N$ (the number of clusters 
normalized to galaxy luminosity $M_V = -15$) than non-nucleated dE, even when
the nuclei are not counted as clusters, which also suggests a different origin
for the two galaxy types (Miller et al. 1998, 2001). 
 And finally, dynamical 
friction calculations suggest that if nuclei are formed
by the merger of decayed clusters, faint dEs should be more
likely to be nucleated because they have shorter cluster decay times for
fixed cluster mass; however 
it is the bright dEs which are more likely to be nucleated (Sandage, Binggeli,
\& Tammann 1985).  An alternative scenario to dynamical friction as the
dE nuclear formation mechanism suggests that dE nuclei are the
remnants of the last burst of star-formation during the transition from 
dwarf irregulars (dIs) to dEs (Davies \& Phillips 1988).

Globular cluster formation is often associated with periods of 
vigorous star formation, such as the initial monolithic gravitational collapse 
of proto-galaxies (Harris 1991)
or starbursts sparked by galaxy mergers and interactions (e.g. the Antennae,
Whitmore \& Schweizer 1995). 
The characteristics of the resulting cluster system record the time
since formation and the properties of the host galaxy.  
The colors of the clusters reflect their ages and abundances, which
constrain the star formation history of the host galaxy.  The current
number of clusters depends on the globular cluster formation efficiency
and the cluster destruction efficiency of the host galaxy, which may vary with
Hubble type and galaxy mass.  The globular cluster specific frequency 
S$_N$ is a function of Hubble type and may
give insight into the formation of the globular cluster system (GCS)
 and host galaxy. The present 
spatial distribution of the clusters  traces the host galaxy's gravitational
potential and dark matter component as well as constrains the initial
radial distribution and kinematics of the cluster population. 

In the Local Group, observations of the dSph GCSs
have presented some interesting puzzles.  
Recent observations of the globular cluster systems of the Local Group
dwarf spheroidals Sagittarius and Fornax have shown a significant age
spread in these systems (Buonanno et al. 1999; Fusi Pecci et al. 1995; 
Montegriffo et al. 1998).  The epoch of globular cluster formation for these
two galaxies lasted for least 3-7 Gyr and the youngest globular cluster 
is $\sim$ 5-7 Gyr old.  These galaxies also have extremely high specific
frequencies  (S$_N > 25$) - 
in other words, they have more clusters than expected for
such faint, low mass systems (van den Bergh 1995). Finally, 
despite their short dynamical friction timescales,
both Fornax and Sagittarius have retained extended globular systems
(Oh, Lin, \& Richer 2000).  

We have embarked on a Hubble Space Telescope (HST) snapshot survey of globular cluster systems
in dE galaxies in nearby clusters and groups.  A primary goal of
the study is to determine whether the globular cluster specific frequency
is high ($S_N \gtrsim 3$), like giant ellipticals, or low ($S_N < 1$),
like dwarf irregulars.  The relatively high values of $S_N$ found in the 
first half of our survey (Miller et al. 1998) suggest that 
dEs (and especially nucleated dEs) are not simply the faded remnants of
dwarf irregular galaxies that have been stripped of their gas. 
In this paper, we consider the constraints on dE evolution provided
by  their nuclear properties and the spatial distributions of their 
globular clusters and search for evidence of dynamical friction processes.

A globular cluster orbiting a galaxy will experience a drag force due
to its gravitational interaction with the surrounding stars and
dark matter.  Over time,
this ``dynamical friction'' (Chandrasekhar 1943) 
will cause the globular cluster to lose
energy and spiral in toward the bottom of the gravitational well.
The timescale over which this happens depends on the mass of the 
globular cluster, its orbit, and the velocity distribution function
 of the particles
with which it is interacting.  In \S 3 we show that for the simplest 
assumptions  -- that the globular clusters began with the same radial
distribution as the stars, that their orbits are isotropic, and that
the dark matter distribution is an isothermal or King sphere  -- the dynamical
friction timescales are significantly shorter than a Hubble time.
In \S 4 we then examine the radial distribution of globular clusters 
in the  51 dE galaxies in our HST snapshot survey for evidence of these
dynamical friction processes. We find that the bright clusters 
(excluding the nuclei) show a 4 $\sigma$
depletion relative to an exponential radial profile, 
consistent with the orbital decay of these clusters into
the dE centers. 
In \S 5 we simulate the formation of dE nuclei via dynamical friction.
Our simulations for the fainter dEs predict nuclei several 
magnitudes brighter than observed.  These dEs have high globular cluster
specific frequencies and extended globular cluster systems but very
short predicted decay timescales.  We discuss in
\S 6 the processes  which may be working against the orbital 
decay of the globular clusters into a single bright nucleus in the 
faintest dEs.

\section{Observations and Data Analysis}

HST WFPC2 snapshot images (HST Cycle 6
6352 and Cycle 7 7377 programs) in F555W (2 $\times$ 230 seconds) and F814W 
(300 seconds)
were taken of each galaxy, with the galaxy centered on chip 3.
The DAOFIND detection algorithm was run on the F555W image and circular
aperture photometry of all detected objects was done using a 3 pixel
radius aperture.  The average aperture corrections for the WF chips are
-0.275 $\pm$ 0.014 for F555W and -0.307 $\pm$ 0.015 for F814W. 
The $V-I$ colors (0.5 $< V-I <$ 1.5) and size (FWHM $<$ 2.5 pixels) 
were used to select globular cluster candidates.
Objects detected on chips 2 and 4 that met our selection criteria 
were considered background/foreground objects and were used to determine the 
background/foreground contamination of chip 3.
All of the dE nuclei in the sample were compact enough to be considered
globular cluster candidates, and, thus, were 
not well resolved even with the high
resolution of HST WFPC2 ($\sim$  0.1 \arcsec\ per pixel ). 
We assumed distance moduli of 31.2 for the Virgo Cluster, 31.4 for the
Fornax Cluster, and 30.3 for the Leo Group.
 More details on the data reduction and object selection can be found in 
Miller et al. (1998). 

 Assuming that the globular clusters have a luminosity
function (LF) similar to the Local Group GCLF, we detect 85-90\% of 
the clusters in Virgo and Fornax (down to $M_V \sim -6$) (Figure 1a) and 98\%
of the clusters in Leo.  Therefore we multiply the total number of Virgo and
Fornax globular cluster candidates by 1.15 to correct for incompleteness.
However, because we wish to obtain the radial profiles of the globular
cluster systems, we must also determine the completeness as a function of
radius from the center of the host galaxy.  Dwarf galaxies are generally
of low surface brightness, $\mu_o >$ 21 V magnitudes/arcsec$^2$ (Binggeli
\& Cameron 1991),  
so we expect the
globular cluster detection efficiency to worsen significantly 
only in the central
regions of the dE.  Simulations of a GCS for a V=15.2 ($M_V =-16$ at Virgo) dE
show that the cluster detection rate is $\sim 70\%$ for
r $< r_0$ and is $> 95\%$ for r $> 3 r_0$ (Figure 1b). It is the
faint clusters which are missed and  clusters brighter than
$M_V= -7.0$ at the distance of Virgo and Fornax are detected at all radii.

Zero to twenty-five globular cluster candidates were found for 
each galaxy (Tables 1 and 2).  The IRAF task 
ELLIPSE was used to determine the center, exponential scalelength
$r_0$, and angle of 
orientation for each galaxy.  The isophotal radial distance (the
semi-major axis length of the intersecting galactic isophote) for each 
globular cluster candidate was 
calculated.  Because each galaxy possesses only a small number of globular 
cluster candidates, a 
composite globular cluster radial distribution was created by scaling each
 globular cluster system by the
host galaxy's exponential scalelength.  Before summing the radial profiles
of the entire dE sample, the profiles were corrected for background/foreground
object contamination by subtracting the estimated number of background objects
per radial bin for each dE. The summed profile was then corrected for
radial incompleteness effects.

\section{Dynamical Friction Timescales}

The observed radial distribution of a galaxy's GCS
depends on 1) the initial radial distribution; 2) the cumulative effects of 
dynamical friction (which in turn depend on the host galaxy's gravitational 
potential well and dynamical structure, as well as the mass distribution of 
the clusters); and 3) the efficiency of destruction mechanisms near the 
galaxy's center such as bulge and disk shocking. Because dEs are  
without bulges or disks, we will assume that the destruction of 
observable globular clusters due such shocks is minimal.
However, if dEs possess central densities $>$ 1 M$_{\odot}$  pc$^{-3}$, globular
clusters passing through the central regions may experience some tidal
stripping. Here we assume that dynamical friction
is the dominant destruction mechanism.

The simplest 
hypothesis for the dE GCSs is that they
started out with the same spatial and velocity distributions as the
underlying stellar population.  Durrell et al. (1996) summed the radial 
distributions for 11 dE GCS and found that dE GCS systems extended no
more than 2.5 kpc from the galaxies' centers and followed the same 
distribution as the stellar light.  However, an earlier study by 
Minniti et al. (1996) of the summed GCS of four Local Group dEs found
evidence for more extended dE GCS.  The GCS of giant galaxies are generally
more spatially extended than the stars; no GCS has been found to be more
centrally concentrated that the rest of the galaxy (Harris 1991).  
Because most GCSs are
associated with a spatially extended component of galaxies, globular clusters 
could have formed before the bulk of the galaxy's star formation
 during a rapid monolithic collapse of
proto-galactic halos.  Another scenario predicting initially extended radial
distributions suggests that globular clusters 
form from cool gas in the outer halo shocked
by a starburst-driven galactic wind (Taniguichi, Trentham, \& Ikeuchi 1999).  
However,
the extended GCSs observed in giant 
ellipticals may result from the preferential
destruction of globular clusters 
near the galaxy's center by bulge and disk shocking (Murali \& Weinberg 1997;
Vesperini 1997).
Globular clusters may also form out of tidally shocked gas during 
galaxy mergers and
strong interactions; observations of forming star clusters in nearby mergers
show that these clusters have the same radial distribution as the
underlying stellar populations (Whitmore et al. 1993; Schweizer et al. 1996).  
For the purposes of this paper, we will
assume that dE GCSs initially had the same spatial distribution as the 
host galaxy's stars.  
 
Constraints on the structure of dE gravitational potential wells are
 particularly 
hard to come by, given the difficulty in observing such faint, generally 
low surface-brightness objects.  Most studies have been limited to using the 
central stellar velocity dispersions ($\sigma$) of Local Group objects.  
Peterson \& Caldwell (1993) find a tight correlation between the dE luminosity 
and $\sigma$, in agreement with supernovae-driven wind dE formation
models (Dekel \& Silk 1986).  Bright dEs with $M_V \sim -17$ have $\sigma$
$\sim$ 60 km s$^{-1}$ and faint dEs with $M_V \sim -12$ 
have $\sigma$ $\sim$ 15 km s$^{-1}$.
Mass estimates from these velocity measurements assuming isotropic
King profiles suggest that dEs are increasingly dominated by dark matter 
at fainter magnitudes (Aaronson 1983; Pryor \& Kormendy 1990). 
However, the stellar velocity dispersion for some dEs are not isotropic - 
measurements of the anisotropy parameter $(v/\sigma)^{*}$ ($= \frac{v}{\sigma}/
\sqrt{\epsilon/(1-\epsilon)}$, where $v$ is the rotational velocity, $\sigma$ is
the mean velocity dispersion, and $\epsilon$ is the ellipticity at
the radius where $v$ is measured)  
yield values less than 0.4 (Bender \& Nieto 1990).  
But the Bender \& Nieto (1990) sample only contains two dEs and for
simplicity we assume that dE dark matter halos are isotropic. 
Here we calculate the dynamical friction timescales for two isotropic 
potentials - the isothermal halo and the King model.  We assume dEs to have
circular velocities ($v_c = \sqrt{2} \sigma$ for the isothermal model,
$v_c = \sqrt{3} \sigma$ for the King model) between 10-100 km s$^{-1}$ for
$M_V = -12$ to $-17$.  

The time for the orbit of a globular cluster of mass M in an isothermal
halo with a circular velocity $v_c$ to decay from an initial radius $r_i$ 
to the center is
\begin{eqnarray}
t_{DF} = \frac{2.64 \times 10^{2}}{\rm{ln} \Lambda} \left( 
 \frac{r_i}{2\ \rm{kpc}}  \right) ^{2}
\left(\frac{v_c}{250\ \rm{km \ s^{-1}}} \right)
 \left(\frac{10^6 \rm{\Msun}}{\rm{M}} \right) \rm{Gyr}
\end{eqnarray}
(Binney \& Tremaine 1987). If we assume that the globular cluster
distribution generally follows the stellar light, the initial globular
cluster radii should be comparable to dE half-light (effective) radii ($\sim$
500 pc to 1.5 kpc, Peterson \& Caldwell 1993).
  
The Coulomb logarithm $\rm{ln} \Lambda$ is given by 
\begin{eqnarray}
\rm{ln} \Lambda = \rm{ln} \left( \frac{b_{max} V^2}{G (M + m)} \right),
\end{eqnarray}
where $b_{max}$ is the maximum impact parameter between the cluster and
interacting particle, V is the typical velocity of the system (for which
we assume V $\sim v_c$) and $m$ is the mass of the particle (star).
The value of $b_{max}$ for dE galaxies is unknown and may range between
0.5 to 3 kpc.  However, this introduces only a 10-15 \% uncertainty 
to the dynamical friction timescale.  We assume $b_{max}$ = 1 kpc for all
of our calculations.
 
Hernandez \& Gilmore (1998) have recently re-addressed the dynamical
friction timescales for dwarf galaxies, assuming dwarf dark matter halos
are well represented by constant density core region of a King distribution
(King 1966).
If the dark matter distribution has an approximately 
constant-density core with radius $r_k$, the dynamical friction timescale
for clusters with $r_i = r_k$ becomes
\begin{eqnarray}
t_{DF}= \frac{4}{3} v_c \left( \frac{r_k}{\rm{kpc}} \right)^2 
\left(\frac{10^5 \rm{\Msun}}{\rm{M}} \right) 
\left( \frac{1}{\rm{ln}\Lambda} \right) \rm{Gyr}.
\end{eqnarray}
(Following Hernandez \& Gilmore, here we have 
defined the ``dynamical friction timescale'' to be the
time for the globular cluster orbit to decay to 1/16th of 
its initial radius.) The core radius
of the King model is defined as $r_k = (9 \sigma^2 / 4 \pi G \rho)^{1/2}$.
The light profiles of dEs are reasonably well fit by King models, although 
bright dEs tend to have central excesses (Binggeli \& Cameron 1991). 
Typical stellar King core radii are on the order of 1 kpc.  
However, dEs are likely to be dark matter
dominated and the extent of their dark matter halos is not well known.
Pryor \& Kormendy (1990) found the best dark matter models for the 
Draco and Ursa Minor dSphs were either isotropic with similar dark matter
and stellar core radii or highly anisotropic with dark matter 
scale radii $\ge$ 10 stellar core radii. 
Dwarf spirals may also be fit with King
profiles, but have larger dark matter core radii (2-8 kpc, Burkert 1995) .

In Figure 2, we plot the dynamical friction timescale as a function
of circular velocity for dwarf-sized isothermal and King halos assuming
different initial/ dark matter core radii and cluster masses.  In
 the isothermal halo,
clusters with masses $\geq$ 3 $\times$ $10^5$ $\Msun$ and initial radii
$\leq$ 1 kpc will spiral into the center in less than 10 Gyr.  In the
King halo, clusters with masses $\geq$ 2 $\times$ $10^5$ $\Msun$ in
a $r_k \leq$ 1 kpc dark matter halo will spiral into the center in less 
than 10 Gyr.  Thus, for simple yet reasonable assumptions, i.e. that dEs
are adequately modeled by isotropic isothermal or King profiles, 
that the dark matter and initial globular cluster distributions
 follow the stellar profile, and that
the  clusters have ages similar to Galactic globular clusters ($>$ 10 Gyr), 
the dE GCSs should be significantly affected by dynamical friction processes.
The effects are stronger for fainter, lower mass dwarf galaxies.

\section{Evolution of dE globular cluster radial distributions}

As the orbits of globular clusters decay under dynamical friction, the massive
clusters will sink into the center more quickly than the smaller clusters.
We have run a simple Monte Carlo simulation to trace the evolution of the
radial distribution of a typical dE globular cluster system. The simulation
uses the analytic expression for radial decay due to dynamical friction
in an isothermal sphere (Equation 1).  
Because dynamical friction decreases energy from
a globular cluster at a rate proportional to $dV^2$ but decreases
angular momentum at a rate proportional to $dV$, a globular cluster
acting under dynamical friction will decay into orbits of maximum angular
momentum (Hernandez \& Gilmore 1998).  Therefore we assume that the orbits
of the globular clusters are circular. 
The clusters initially follow the exponential profile of the underlying
stellar population and have the present day globular cluster mass function,
independent of radius.
The simulation was run 100,000 times for three cases:
$v_c =$ 50 km s$^{-1}$ and $r_0 =$ 2 kpc;  $v_c =$30 km s$^{-1}$ and $r_0 =$ 1 kpc; 
and $v_c= $15 km s$^{-1}$ and $r_0 = $ 0.5 kpc (Figure 3).

The simulated globular cluster radial profiles in Figure 3 demonstrate
the increasing impact of dynamical friction on smaller, lower mass dwarfs.
For the first two cases, dynamical friction has little effect on the summed
profile of the entire GCS and only the massive clusters within the inner
scalelengths fall into the galaxy's core within 10 Gyr. In the 50 km s$^{-1}$ dE,
 all clusters with masses $> 5 \times 10 ^5 \Msun$ and  
radii $<$ 1 kpc (1/2 $r_0$) merge into the center within 10 Gyr.  
In the 30 km s$^{-1}$ dE,  all clusters with masses $> 5 \times 10 ^5 \Msun$ 
and initial radii $<$ 2 kpc (2 $r_0$) merge into the center within 10 Gyr.  
As the inner-most massive clusters are destroyed, some of the massive clusters
originally found in the outer regions sink into the central few scalelengths.
The remaining less massive and more extended
clusters sink more slowly inwards and retain the exponential radial
distribution in the outer regions of the dE.  
However, in the 15 km s$^{-1}$, 0.5 kpc simulation, after 10 Gyr  $\sim$ 85\% of 
all clusters have merged into the center and  99.9\% of $M_V < -8.0$ 
clusters have
been destroyed. In the faintest and smallest dE, most of the globular
clusters should merge to form a bright nucleus in less than a Hubble time.

We have examined the summed radial distribution of both the bright 
($M_{V} < -8$), non-nucleus cluster candidates and the entire GCSs
of our dE sample for evidence of these dynamical friction processes.
Because our sample includes dEs with $-11.9 > M_B > -17.3$,
we expect the impact of dynamical friction on our summed radial distribution
will be most noticeable in the deficit of massive clusters in the 
inner regions of dE.
For each dE, we typically detect only a few globular 
cluster candidates in excess of the background. Without radial velocities, 
it is impossible to determine if a given
globular cluster candidate is associated with the dE or is a background
galaxy or foreground star. Figure 4 shows the magnitude distribution of 
globular cluster candidates and background/foreground
 objects satisfying the same 
color and compactness criteria. Significant contamination by background 
objects affects the brightest bins; roughly 30\% of the brightest 
cluster candidates which fall on chip 3 
and have $M_{V} < -8$ are likely to be background objects. 
Assuming $M/L_V$ = 2, a $M_V = -8$ cluster has a
 mass $\sim$ $2.5 \times 10^5 \Msun$.  
The dynamical friction timescale for a
$2.5 \times 10^5 \Msun$ cluster 
with an initial radius of 1 kpc in an isothermal halo
with $v_c \sim$ 40 km s$^{-1}$ is $\sim$ 5 Gyr, thus the dynamical friction
effects should be apparent for these bright clusters.

We have scaled the entire 
globular cluster system of each dE by the host galaxy's exponential 
scalelength $r_0$, subtracted out the background contamination,
 and summed the radial distributions of our 51 dEs to 
create a composite globular cluster radial distribution (top half of Figure 5).
The nuclei are excluded from this radial distribution.
The summed distribution was corrected for radial dependence on incompleteness,
as determined in Figure 1b.
This composite globular cluster radial profile is well fit by 
an exponential profile and follows the light profile of dEs:
\begin{equation}
\rm{ln} (N_{gc}/area) = (-0.97 \pm 0.12) \times (\it{\frac{r}{r_0}}) 
\rm{\ +\ (4.76 \pm 0.45) \ (RMS=0.29).}
\end{equation}
No deficit below the exponential
profile for the entire summed globular cluster distribution is apparent;
however, the effect of dynamical friction on the total distribution
is likely to be undetectable (Figure 3).

In the bottom of Figure 5, we have plotted the radial distribution of the 
bright ($M_V < -8.0$) non-nuclear cluster candidates,
corrected for the 30\% contamination,
scaled by the exponential scalelength of the galaxy and summed over the
entire dE sample. We have done no correction for incompleteness because
the correction should be very small for clusters brighter than $M_V = -7.0$,
even in the central regions of the dEs.
The best linear fit to this profile is $\sim 4 \sigma$ shallower than
an exponential profile:
\begin{equation}
\rm{ln} (N_{gc}/area) = (-0.68 \pm 0.07) \times (\it{\frac{r}{r_0}})
\rm{\ +\ (2.93 \pm 0.25) \  (RMS =0.21).}
\end{equation}
Thus, it appears that there is a significant deficit of bright, massive
globular clusters in the inner regions of our sample dEs.

\section{Formation of dE nuclei via dynamical friction}

Our Monte Carlo simulation of the evolution of dE globular cluster systems'
radial profiles shows that massive clusters quickly spiral into the centers
of dE. 
The brightest globular cluster candidate in a dE is often the nucleus with 
$<M_V> = -10.3$ (M $\sim$ 3 $\times$ 10$^6$ $\Msun$, assuming $M/L_V$ =2).
Most of the nuclei in our sample lie in a fairly small $(V-I)$ range between
0.85 and 1.10, consistent with a 12 Gyr stellar 
population and a metallicity range between 1/100 and 1/5 solar (Figure 6).  
 The fainter nuclei are
also found in fainter galaxies, which generally have fewer clusters overall
(top of Figure 7). There is a slight color-magnitude
relation where the fainter nuclei are bluer.
One can imagine a scenario where fainter, less massive galaxies initially 
have fewer clusters and lower metallicities, and form fainter, bluer nuclei.
The nuclei of FCC25 and FCC1714 lie outside of these trends. FCC 25 is $\sim$ 
0.4 magnitudes redder than the other nuclei and could
be a background galaxy.  FCC 1714 is $\sim$ 0.3 magnitudes bluer and could
also be a background/foreground object or a $<$ 1 Gyr old cluster.
 
We ran another set of Monte Carlo simulations
 to predict the magnitude of each dE
nucleus in our sample of dE,N based on each galaxy's intrinsic 
properties (Table 1).  Again, the simulation randomly sampled 
the cluster mass function
and an exponential radial distribution and allowed the clusters to
orbitally decay for
5 Gyr.  Any cluster with a final radius $\leq$ 0 was merged into the nucleus.
The velocity dispersion $\sigma$ was
estimated using the $\sigma$ - luminosity relation for dEs (Dekel \& Silk 1987,
Peterson \& Caldwell 1993) and converted to circular velocity assuming
$v_c = \sqrt{2} \sigma$.  Each galaxy's exponential scalelength was measured
from our HST WFPC2 I image (Table 1).
The initial number of clusters was estimated by correcting the present
day number of clusters by the dynamical friction destruction efficiency.
The globular cluster destruction efficiency due to dynamical friction 
depends on the host galaxy's scalelength and circular velocity.  
dE,N with less than 2 detected clusters were excluded due to the uncertainty
in calculating the globular cluster destruction efficiency.
After 5 Gyr,
 $\sim$ 15\% of globular clusters will merge into a nucleus in a $v_c =$
50 km s$^{-1}$, $r_0 =$ 2 kpc dE and $\sim$ 80\% of globular clusters will merge 
into a nucleus in a $v_c=$ 15 km s$^{-1}$, $r_0 =$ 500 pc dE. 
 Note that, if dynamical friction is as efficient at destroying 
globular clusters as our simulations suggest, then
the faintest dEs initially formed with up to ten times as many clusters as
we observe today.

Assuming nuclei form via dynamical friction, 
we predict a much weaker trend of nuclear magnitude
with galaxy magnitude than observed (bottom of Figure 7). While the fainter 
galaxies generally have fewer
globular clusters from which to form nuclei, they should have shorter dynamical
friction timescales and thus should be more efficient at destroying clusters
and forming nuclei.  
However, our simulation over-predicts the mass of the nuclei in most of our
sample.
Figure 8 plots the predicted $M_V$ magnitude of the nucleus vs. the observed 
$M_V$ magnitude for each dE,N. 
Nuclei which fall on or below the solid line may have formed via dynamical
friction processes within 5 Gyr.  Thus, even the bright nucleus of VCC1254
($M_V = -12.59 \pm 0.02$) may have been formed from the decayed remnants of 
massive globular clusters.  However, many of the faint nuclei (found 
primarily in the $M_V > -14$ galaxies) are several magnitudes fainter 
than predicted.

Many of these faint galaxies have extremely high $S_N$ (Tables 1 and 2) and
possess extended GCS.  Dynamical friction arguments would suggest
that most of the clusters in these diffuse galaxies should have merged into
a single bright nucleus and the globular cluster 
population we observe today is less than a third of the original population.
Clearly, dynamical friction is not as effective at destroying globular 
clusters and forming nuclei in the faint dEs as our simple simulations predict.
While the faint nuclei may have been formed
via the orbital decay of massive globular clusters,
some mechanism seems to have minimized the dynamical decay of
 the GCSs in these dEs.  

\section{Possible mechanisms working against dynamical friction}
We have detected evidence for the action of dynamical friction
in the radial distribution of the bright non-nuclear globular clusters in
dE galaxies
and found that the brightest nuclei are consistent with formation
via the orbital decay of massive clusters in the center of the host dE.
However, our Monte Carlo simulations of the formation of dE nuclei do not
adequately reproduce the strong observed trend of decreasing nuclear 
luminosity with decreasing
galaxy luminosity.  In the faint dEs, 
some mechanism is working against dynamical friction to prevent the
collapse of the entire globular cluster system.
This result is reminiscent of the observations of the Fornax and 
Sagittarius dSph GCSs.  Both of these $M_V > -14$ Local Group
dSph possess high specific frequencies and extended GCSs, 
despite extremely short dynamical friction timescales.  
In this section we consider the mechanisms which may counteract or 
minimize the dynamical drag on the globular clusters in faint dE.

1) The globular clusters in the faintest dEs could
 be younger than their dynamical friction 
timescales ($\leq$ 5 Gyr).  Many Local Group dSph  possess significant 
numbers of intermediate age stars (1-8 Gyr), and several ``young'' globular 
clusters (4-8 Gyr) are associated with the Sagittarius and Fornax dSph
(Grebel 1997, van den Bergh 2000).
Combined with 
the existence of numerous faint blue galaxies in the field, this suggests that
many dwarf galaxies experienced a secondary starburst between z $\sim$
0.5 - 1.5.  Although the dE globular clusters have colors and luminosities 
similar to the 12-14 Gyr globular clusters in our Galaxy, the age-metallicity
degeneracy prevents us from constraining their ages to better than
a few Gyr.  However, for the Fornax dSph, where globular cluster ages can be
constrained by color magnitude diagrams, the youngest globular 
cluster is still several Gyr older than its dynamical friction timescale
(Oh, Lin, \& Richer 2000). 

2) dEs may have undergone significant mass loss due to 
supernovae-driven winds during the formation of their first stars.
The fractional mass lost increases with decreasing mass, thus the smallest
objects suffered the greatest mass loss (Dekel \& Silk 1986). 
The galaxy's globular cluster system may have become
 more extended than the remaining stars
as a result of the supernovae-driven wind, and take longer to 
spiral inward under the influence of dynamical friction. 
If they formed in the galaxy's halo
before the onset of the wind (McLaughlin 1999) or 
as the wind shocked the
galaxy's outer halo (Taniguichi et al. 1999), they would initially possess 
a radial profile more extended than the stellar light.  The initial
extent of this globular cluster halo may increase with increasing mass loss
and decreasing initial mass. However, it is difficult to see how 
the GCSs could then evolve to 
match the exponential stellar profile at the present day.

3) The structure of the dark matter halos may be significantly
different from the stellar halos.  An increase in either the
dark matter density or the dark matter core radius relative to the
stellar density or core radius will increase the velocity dispersion
of the dark matter and the dynamical friction timescale of the 
globular clusters (Hernandez \& Gilmore 1998). 
The $M/L$ estimates for dEs predict that most of
the mass is in dark matter; thus the interaction with the dark
matter rather than with the stars will dominate 
the drag on the globular clusters.
 The faintest dEs may have high dark matter densities (Dekel \& Silk 1986;
Peterson \& Caldwell 1993)
which would slow down the orbital decay of the clusters.
The Local Group dSph show a trend towards increasing mass to light ratios
with decreasing luminosity (Peterson \& Caldwell 1993; Mateo 1998).
Also, the distribution of dark matter in dEs is 
very poorly constrained. Dwarf spirals and other LSBs for which HI rotation
curves may be measured show dark matter halos which extend several scalelengths
beyond the stellar profiles.  If dEs are gas-stripped dwarf irregulars and
spirals, they too should have extended dark matter halos. 

Expressing the dynamical friction
time scale of Equation (3) in terms of the density (see also Carollo 1999
for a different derivation) yields for clusters starting
with the core radius the relation
\begin{equation} 
t_{DF} = \frac{1.26 \times
10^2}{\ln{\Lambda}}\left(\frac{r_k}{2\ \rm{kpc}}\right)^3 \left(\frac{10^6
\rm{M_\odot}}{\rm{M}}\right)
\left(\frac{\rho}{\rm{M_\odot pc}^{-3}}\right)^{1/2}{\rm{Gyr}}, 
\end{equation}
with the density given by $\rho_0 = 0.55\ (\frac{V_0} 
{100 \ {\rm km \ s^{-1}}})^2 \
(\frac{1 {\rm kpc}}{r_k})^2 \ {\rm M}_\odot\ {\rm pc}^{-3}$.  Our results would
imply dE dark matter densities in excess of $\sim 1$ M$_\odot$ pc$^{-3}$
or dE dark matter core radii $>$ 2 kpc for the faintest dE.

Modified Newtonian Dynamics (MOND) offers a possible alternative to 
dark-matter dominated dEs.  MOND has been successful at explaining the velocity
rotation curves of low surface brightness disk galaxies; dSph and dEs should
also provide a strong test for MOND (McGaugh \& de Blok 1998).  
However, a full derivation of dynamical
friction taking into account the effects of MOND is needed to determine if
MOND is consistent with our results and that is beyond the scope of this paper.

4) The clusters may be heated in such a way that counters their orbital
decay via dynamical friction.  Oh, Lin \& Richer (2000) recently attempted
to identify possible heating mechanisms for  the Fornax dSph GCS.  One
possible mechanism for injecting kinetic energy into the GCS is by scattering
of the globular clusters after close encounters with massive black 
holes ($M_{BH} \geq M_{GC}$).  Such scattering may also occur once a massive
nucleus has formed. However, repeated encounters are likely to
 destroy the clusters.

Oh et al. (2000) also examine the effects of the tidal disruption of the Fornax
dSph by the Galaxy.  While the tidal forces increase the
globular cluster orbital eccentricities and velocity dispersions, the strong 
tidal forces
can not increase the GCS kinetic energy enough to counter the dynamical 
friction drag.  But, if Fornax has experienced significant mass loss 
(50-90 \%) via
tidal stripping over many internal dynamical timescales, the clusters will 
retain their energy in a shallower gravitational potential well and their
orbital radii will increase, countering the drag of dynamical friction.

The dEs in our sample were selected to not be close companions of giant
galaxies, for fear that the dE globular cluster candidates would be
contaminated by those of the giant companion.  However, the tidal
field of the entire cluster of galaxies may play a role in heating
their GCSs.  Oh \& Lin (2000)
model the effects of the extra-galactic tidal field of the Virgo Cluster
on dE GCSs.  dEs within the inner regions of Virgo experience a relatively
weak external tidal perturbation and therefore their GCSs are able to
decay under dynamical friction and coalesce into nuclei.  But in the outskirts
of Virgo, dEs may be tidally disrupted and suffer significant mass loss as
in the Fornax dSph simulation.  Less massive galaxies are more likely to
be disrupted and thus less likely to form nuclei.  

\section{Conclusions}
We have examined high resolution HST WFPC2 V and I images of dE galaxies for
evidence of dynamical friction in their globular cluster systems and 
nuclear properties.  
We find that, within the inner few scalelengths, our sample appears
to be depleted of bright clusters, but there are no other
observable effects on the summed profile.
Monte Carlo simulations of the evolution of dE globular cluster systems
show that the brighter nuclei may have formed via orbital decay 
into the galaxy's core. However, the observed trend of fainter nuclei
in fainter galaxies is much stronger than expected if the nuclei form
via simple dynamical friction processes.  In these galaxies, some mechanism
must be working against the orbital decay of the globular clusters.
Mass loss via supernova-driven winds, high dark matter densities
and/or large dark matter cores, the
formation of new star clusters, and tidal stripping 
may all conspire together to counteract the dynamical drag on the faintest 
dE GCSs and inhibit the formation of bright nuclei.
We will obtain Cycle 9 HST WFPC2 observations of $\sim$ 30 brighter
dEs with larger GCS to improve our statistics.

We wish to thank Rosie Wyse and Brad Whitmore for useful discussions
about dynamical friction and globular cluster detection. 
We also thank our referee, Tad Pryor, for his very constructive and helpful
comments. Support for this work was provided by NASA through grants
GO-06352.01-95A and GO-07377-01-96A from the Space Telescope
Science Institute, which is operated by the Association of Universities
for Research in Astronomy, Incorporated, under NASA contracted NAS5-26555.

\clearpage
\begin{deluxetable}{cccccrrrrrr}
\tabletypesize{\scriptsize}
%\rotate
\setlength{\tabcolsep}{0.05in}
\tablecaption{Nucleated Dwarf Elliptical Galaxies \tablenotemark{1}}
\tablehead{ \colhead{Galaxy} &\colhead{Type} &\colhead{M$_B$ \tablenotemark{2}}  & \colhead{M$_V$} &\colhead{Cycle} &\colhead{M$_{Vnuc}$} &\colhead{V-I$_{nuc}$}
& \colhead{r$_0$(\arcsec)} &\colhead{r$_0$(kpc)}  &\colhead{N$_{gc}$} &\colhead{S$_N$} \tablenotemark{3} }
\startdata
VCC1073  &dE3,N  & -17.0 & -17.4    &  6 & -11.62 $\pm$  0.03 & 1.08 $\pm$  0.04 &    8.6 $\pm$  0.5  & 0.71 $\pm$ 0.04 & 18.8 $\pm$  6.2 &   2.1 $\pm$   0.7\\
FCC136   &dE2,N  & -16.6 & -17.2    &  6 & -11.50 $\pm$  0.03 & 1.07 $\pm$  0.04 &    8.4 $\pm$  1.3  & 0.77 $\pm$ 0.12 & 20.1 $\pm$  5.6 &   2.7 $\pm$  0.8\\
VCC1876  &dE5,N  & -16.3 & -16.6    &  6 & -10.74 $\pm$  0.03 & 1.00 $\pm$  0.04 &   10.5 $\pm$  0.3  & 0.86 $\pm$ 0.03 & 24.7 $\pm$  6.8 &   5.5 $\pm$  1.6\\
VCC1254  &dE0,N  & -16.2 & -16.5    &  6 & -12.59 $\pm$  0.02 & 1.04 $\pm$  0.04 &    7.7 $\pm$  0.2  & 0.63 $\pm$ 0.02 & 24.6 $\pm$  8.2 &   6.2 $\pm$  2.2\\
FCC324   &dS01(8)& -16.1 & -16.4    &  6 &  -8.95 $\pm$  0.04 & 0.91 $\pm$  0.06 &   13.9 $\pm$  0.3  & 1.30 $\pm$ 0.04 & 12.0 $\pm$  4.8 &   3.3 $\pm$  1.4\\
FCC150   &dE4,N  & -15.7 & -16.2    &  6 & -11.11 $\pm$  0.03 & 1.05 $\pm$  0.04 &    5.5 $\pm$  0.3  & 0.51 $\pm$ 0.03 &  7.4 $\pm$  4.2 &   2.4 $\pm$  1.4\\
VCC452   &dE4,N  & -15.4 & -15.8    &  6 &  -8.85 $\pm$  0.04 & 0.94 $\pm$  0.06 &    7.8 $\pm$  0.4  & 0.64 $\pm$ 0.04 & 11.4 $\pm$  5.8 &   5.3 $\pm$  2.7\\
FCC316   &dE3,N  & -15.1 & -15.2    &  6 &  -9.62 $\pm$  0.03 & 1.08 $\pm$  0.05 &    8.2 $\pm$  0.3  & 0.75 $\pm$ 0.03 & 17.8 $\pm$  5.7 &  14.6 $\pm$  4.9\\
FCC174   &dE1,N  & -14.7 & -15.3    &  6 &  -9.90 $\pm$  0.03 & 0.93 $\pm$  0.05 &    4.9 $\pm$  0.5  & 0.45 $\pm$ 0.05 &  9.8 $\pm$  3.8 &   7.2 $\pm$  2.9\\
VCC503   &dE3,N  & -14.4 & -14.5    &  6 &  -9.43 $\pm$  0.03 & 0.89 $\pm$  0.05 &    6.3 $\pm$  0.4  & 0.52 $\pm$ 0.03 &  1.7 $\pm$  3.6 &   2.8 $\pm$  6.0\\
FCC254   &dE0,N  & -13.8 & -14.7    &  6 & -10.70 $\pm$  0.03 & 0.92 $\pm$  0.04 &   10.5 $\pm$  0.5  & 0.97 $\pm$ 0.05 &  5.7 $\pm$  4.1 &   7.7 $\pm$  5.7\\
FCC25    &dE0,N  & -13.7 & \nodata  &  7 &  -9.80 $\pm$  0.03 & 1.37 $\pm$  0.04 &    6.1 $\pm$  0.3  & 0.57 $\pm$ 0.03 &  5.2 $\pm$  3.7 &   8.2 $\pm$  5.8\\
VCC896   &dE3,N  & -13.4 & \nodata  &  7 &  -8.60 $\pm$  0.04 & 1.01 $\pm$  0.09 &    8.8 $\pm$  0.6  & 0.75 $\pm$ 0.05 &  5.8 $\pm$  3.9 & 12.0 $\pm$  8.1\\
LGC50    &dE,N   & -13.3 & -14.0    &  6 &  -8.90 $\pm$  0.03 & 0.82 $\pm$  0.05 &   10.2 $\pm$  0.4  & 0.54 $\pm$ 0.02 &  5.5 $\pm$  3.1 &  13.9 $\pm$  9.2\\
VCC240   &dE2,N  & -13.0 & -13.9    &  6 &  -9.89 $\pm$  0.03 & 0.91 $\pm$  0.04 &    9.5 $\pm$  1.0  & 0.78 $\pm$ 0.08 &  8.6 $\pm$  4.1 &  23.6 $\pm$  14.\\
VCC529   &dE4,N  & -13.0 & \nodata  &  7 &  -9.85 $\pm$  0.03 & 0.92 $\pm$  0.04 &    6.6 $\pm$  0.5  & 0.56 $\pm$ 0.05 &  3.5 $\pm$  3.6 & 10.4 $\pm$ 10.9\\
VCC1530  &dE2,N  & -12.9 & \nodata  &  7 &  -9.90 $\pm$  0.03 & 0.88 $\pm$  0.04 &    5.9 $\pm$  0.3  & 0.50 $\pm$ 0.03 &  7.5 $\pm$  4.2 & 28.6 $\pm$ 13.5\\
VCC1783  &dE5,N  & -12.8 & \nodata  &  7 &  -9.15 $\pm$  0.03 & 0.95 $\pm$  0.05 &   11.6 $\pm$  0.4  & 0.95 $\pm$ 0.03 &   0  $\pm$  3.5 &   0 $\pm$ 13.9	  \\
VCC1714  &dE4,N  & -12.7 & \nodata  &  7 & -12.03 $\pm$  0.02 & 0.58 $\pm$  0.03 &   12.0 $\pm$  0.8  & 1.02 $\pm$ 0.07 &  0.6 $\pm$  3.4 &  2.3 $\pm$ 13.5\\
VCC1272  &dE1,N  & -12.7 & \nodata  &  7 &  -7.84 $\pm$  0.06 & 0.87 $\pm$  0.09 &    4.2 $\pm$  0.4  & 0.35 $\pm$ 0.04 &  1.7 $\pm$  3.5 &  6.9 $\pm$ 14.1\\
FCC238   &dE5,N  & -12.7 & \nodata  &  7 &  -9.27 $\pm$  0.05 & 0.91 $\pm$  0.04 &    7.0 $\pm$  0.7  & 0.65 $\pm$ 0.07 &  2.3 $\pm$  2.0 &   9.2 $\pm$	8.0\\
FCC189   &dE4,N  & -12.6 & \nodata  &  7 &  -9.12 $\pm$  0.03 & 0.83 $\pm$  0.05 &    2.4 $\pm$  0.3  & 0.22 $\pm$ 0.03 &  1.2 $\pm$  3.3 &   5.5 $\pm$ 15.9\\
VCC1252  &dE0,N  & -12.4 & \nodata  &  7 &  -9.21 $\pm$  0.03 & 0.90 $\pm$  0.05 &    4.7 $\pm$  0.5  & 0.39 $\pm$ 0.04 &  6.3 $\pm$  3.8& 33.2 $\pm$  20.0\\
VCC1363  &dE3,N  & -12.2 & \nodata  &  7 &  -8.98 $\pm$  0.03 & 0.93 $\pm$  0.05 &    3.7 $\pm$  0.2  & 0.30 $\pm$ 0.02 &   0 $\pm$  2.2 &   0 $\pm$ 15.2	  \\
VCC1077  &dE0,N  & -12.0 & \nodata  &  7 &  -8.69 $\pm$  0.04 & 0.94 $\pm$  0.06 &    3.3 $\pm$  0.3  & 0.28 $\pm$ 0.03 &  1.2 $\pm$  3.9 &  8.7 $\pm$ 29.4\\
FCC59    &dE0,N  & -12.0 & \nodata  &  7 &  -8.69 $\pm$  0.04 & 0.85 $\pm$  0.06 &   11.0 $\pm$  2.2  & 1.02 $\pm$ 0.20 &   0 $\pm$  3.1 &   0 $\pm$ 25.8	  \\
FCC146   &dE4,N  & -11.9 & \nodata  &  7 &  -8.06 $\pm$  0.05 & 1.04 $\pm$  0.08 &    3.3 $\pm$  0.4  & 0.31 $\pm$ 0.04 &  1.7 $\pm$  2.9 & 15.7 $\pm$ 26.6\\
\enddata
\tablenotetext{1}{ Cycle 6 data also presented in Miller et al. 1998.}
\tablenotetext{2}{ from Ferguson \& Sandage 1990}
\tablenotetext{3}{$S_N = N_{gc} \times 10^{0.4(M_V + 15.0)}$, $M_V$ assumed
to be $M_B - 0.8$ where $M_V$ not measured.}
\end{deluxetable}

\clearpage
\begin{deluxetable}{cccccrrrr}
\tabletypesize{\scriptsize}
%\rotate
%\setlength{\tabcolsep}{0.04in}
\tablecaption{Non-nucleated Dwarf Elliptical Galaxies \tablenotemark{1}}
\tablehead{ \colhead{Galaxy} &\colhead{Type} &\colhead{M$_B$ \tablenotemark{2}}  & \colhead{M$_V$} &\colhead{Cycle} 
& \colhead{r$_0$(\arcsec)} &\colhead{r$_0$(kpc)}  &\colhead{N$_{gc}$} &\colhead{S$_N$}\tablenotemark{3} }
\startdata
VCC9     &dE1,N? & -17.3 & -17.5    &  6   & 18.3  $\pm$  0.6  & 1.54 $\pm$ 0.05 & 22.9 $\pm$  6.3  &	2.3 $\pm$ 0.7 \\
VCC917   &dE6	 & -16.3 & -16.3    & 6    &  6.0  $\pm$  0.2  & 0.50 $\pm$ 0.01 &  6.8 $\pm$  5.3  &	2.1 $\pm$ 1.6\\
VCC118   &dE3	 & -15.6 & -15.3    & 6    &  8.0  $\pm$  0.6  & 0.67 $\pm$ 0.05 &  3.4 $\pm$  3.7  &	2.5 $\pm$ 2.7\\
VCC1577  &dE4	 & -15.4 & -15.8    & 6    &  7.1  $\pm$  0.4  & 0.59 $\pm$ 0.03 & 14.9 $\pm$  5.5  &	7.3 $\pm$ 2.8\\
LGC47    &dE	 & -15.1 & -15.5    & 6    & 12.9  $\pm$  0.6  & 0.68 $\pm$ 0.03 &  3.5 $\pm$  3.9  &	2.2 $\pm$ 2.4\\
VCC1762  &dE6	 & -15.0 & -15.2    & 6    &  6.9  $\pm$  0.2  & 0.58 $\pm$ 0.02 &  4.0 $\pm$  3.7  &	3.4 $\pm$ 3.1\\
FCC110   &dE4	 & -14.6 & -14.9    & 6    & 10.8  $\pm$  0.4  & 0.99 $\pm$ 0.03 &   0 $\pm$  3.8  &	 0 $\pm$ 4.2\\
FCC48    &dE3	 & -14.3 & -14.8    & 6    &  6.9  $\pm$  0.4  & 0.63 $\pm$ 0.04 &  5.1 $\pm$  4.4  &	6.1 $\pm$ 5.3\\
VCC1651  &dE5	 & -14.2 & -14.6    & 6    & 19.2  $\pm$  1.1  & 1.61 $\pm$ 0.09 &  1.1 $\pm$  3.9  &	1.6 $\pm$ 5.7\\
FCC64    &dE5	 & -13.9 & -14.2    & 6    &  8.3  $\pm$  0.7  & 0.77 $\pm$ 0.06 &   0 $\pm$  2.6  &	 0 $\pm$ 5.4\\
FCC212   &dE1?   & -13.8 & \nodata  & 7    & 15.1  $\pm$  0.8  & 1.39 $\pm$ 0.07 &  6.9 $\pm$  4.2  &  10.0 $\pm$ 6.1\\
FCC242   &dE5	 & -13.6 & \nodata  & 7    &  9.3  $\pm$  0.6  & 0.85 $\pm$ 0.05 &   0 $\pm$  2.2  &   0 $\pm$ 4.2  \\
VCC1729  &dE5?   & -13.4 & \nodata  & 7    &  6.3  $\pm$  0.4  & 0.53 $\pm$ 0.03 &  1.2 $\pm$  3.0  &	2.4 $\pm$ 6.3\\
VCC2029  &dE3	 & -13.0 & -13.9    & 6    &  5.0  $\pm$  0.3  & 0.42 $\pm$ 0.02 &  1.1 $\pm$  3.6  &	3.1 $\pm$10.4\\
FCC218   &dE4	 & -12.9 & \nodata  & 7    &  5.6  $\pm$  0.5  & 0.51 $\pm$ 0.05 &   0 $\pm$ 2.6  &	0 $\pm$ 9.4\\
VCC996   &dE5	 & -12.8 & \nodata  & 7    &  7.0  $\pm$  0.4  & 0.59 $\pm$ 0.03 &  1.7 $\pm$ 3.2  &    6.2 $\pm$11.8\\
VCC1877  &dE2	 & -12.6 & \nodata  & 7    &  7.4  $\pm$  0.2  & 0.62 $\pm$ 0.02 &  4.0 $\pm$ 3.2  &   17.6 $\pm$14.1\\
FCC304   &dE1	 & -12.6 & \nodata  & 7    &  7.2  $\pm$  0.9  & 0.66 $\pm$ 0.09 &   0 $\pm$ 2.8  &	0 $\pm$ 13.4 \\
VCC1781  &dE4	 & -12.5 & \nodata  & 7    &  5.6  $\pm$  0.4  & 0.47 $\pm$ 0.04 &  4.0 $\pm$ 3.2  &   19.3 $\pm$15.5\\
VCC646   &dE3	 & -12.4 & \nodata  & 7    &  5.6  $\pm$  0.2  & 0.47 $\pm$ 0.02 &  6.3 $\pm$ 3.8  &   33.1 $\pm$ 20.\\
FCC246   &dE2	 & -12.3 & \nodata  & 7    &  6.6  $\pm$  0.7  & 0.61 $\pm$ 0.07 &   0 $\pm$ 2.6  &   0 $\pm$ 13.4\\
FCC144   &dE0	 & -12.2 & \nodata  & 7    &  3.7  $\pm$  0.4  & 0.34 $\pm$ 0.04 &  0.6 $\pm$ 3.4  &    3.6 $\pm$ 21.4\\
FCC27    &dE2	 & -12.1 & \nodata  & 7    &  3.6  $\pm$  0.5  & 0.33 $\pm$ 0.04 &   0 $\pm$ 2.8  &	0 $\pm$ 21.2 \\
VCC607   &dE7	 & -12.0 & \nodata  & 7    &  7.2  $\pm$  0.7  & 0.60 $\pm$ 0.06 &   0 $\pm$ 2.7  &	0 $\pm$ 22.5 \\
\enddata
\tablenotetext{1}{ Cycle 6 data also presented in Miller et al. 1998.}
\tablenotetext{2}{ from Ferguson \& Sandage 1990}
\tablenotetext{3}{$S_N = N_{gc} \times 10^{0.4(M_V + 15.0)}$, $M_V$ assumed
to be $M_B - 0.8$ where $M_V$ not measured.}
\end{deluxetable}

\clearpage
\begin{figure}
\plottwo{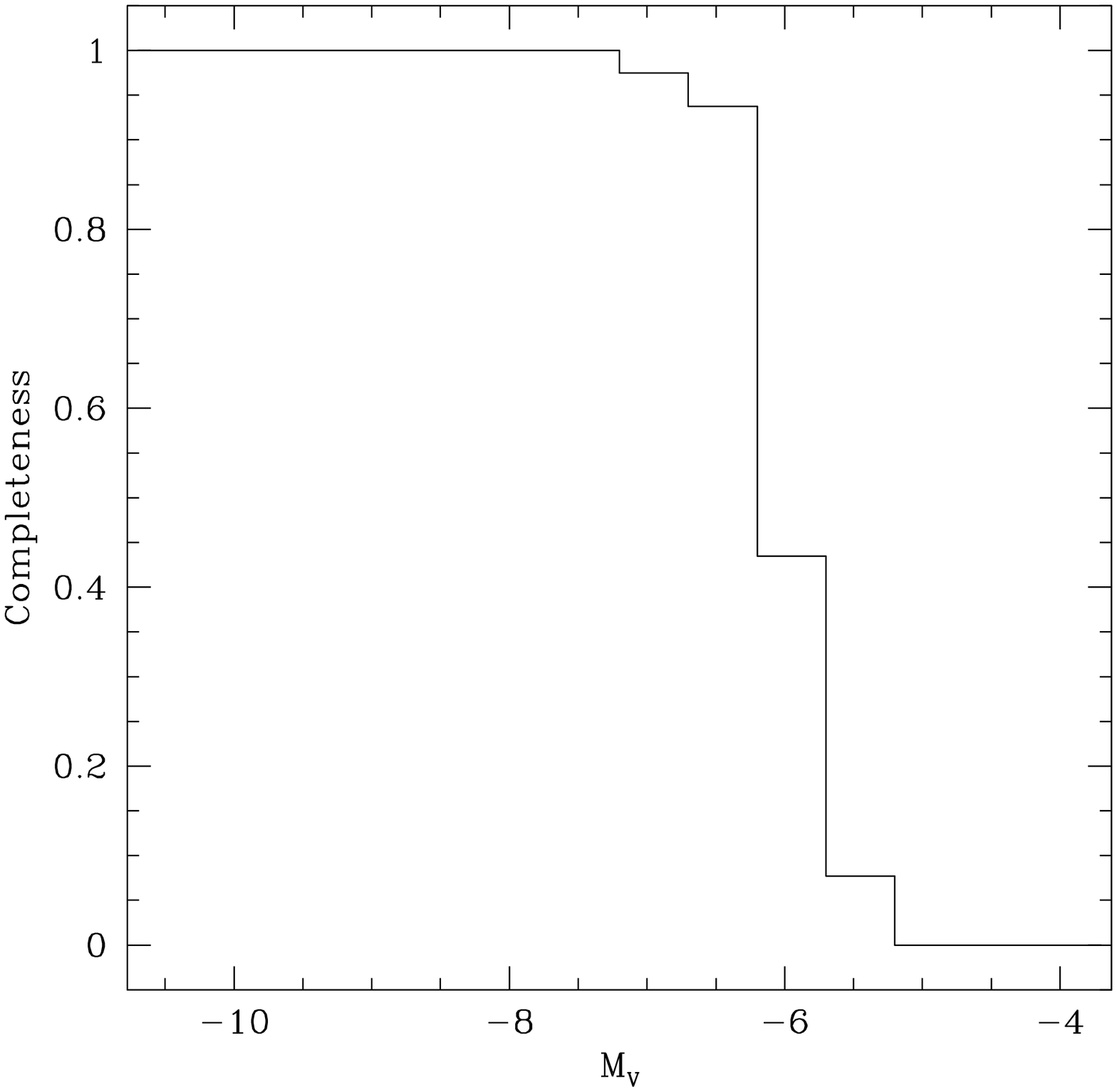}{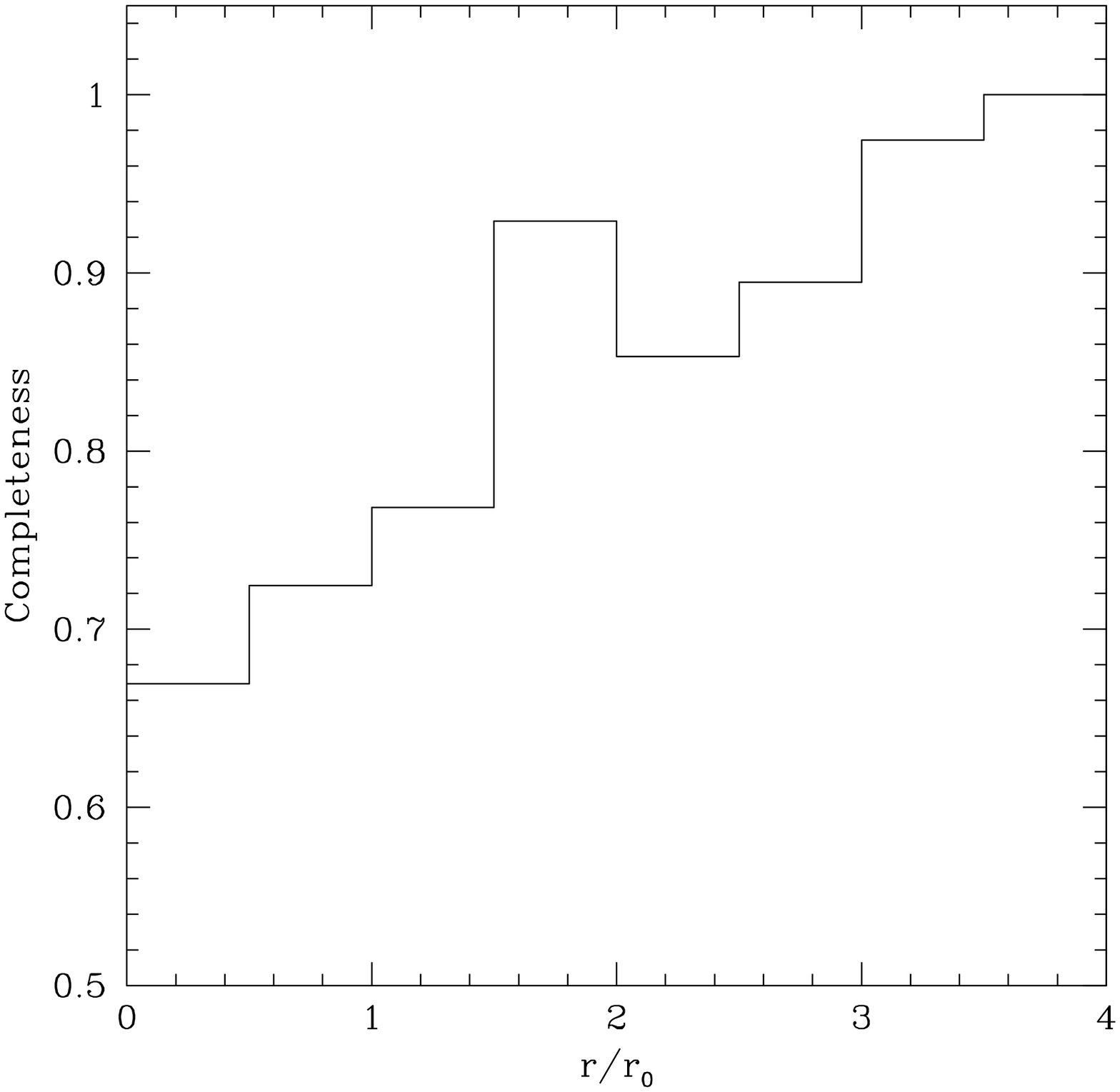}
\caption{Completeness of the globular cluster detection as a function
of globular cluster luminosity and distance from host galaxy center for a $V$ = 15.2, $r_0$
= 15 \arcsec\ dE ($M_V$ = -16.0 at the distance of Virgo).}
\end{figure}

\clearpage
\begin{figure}
\plotone{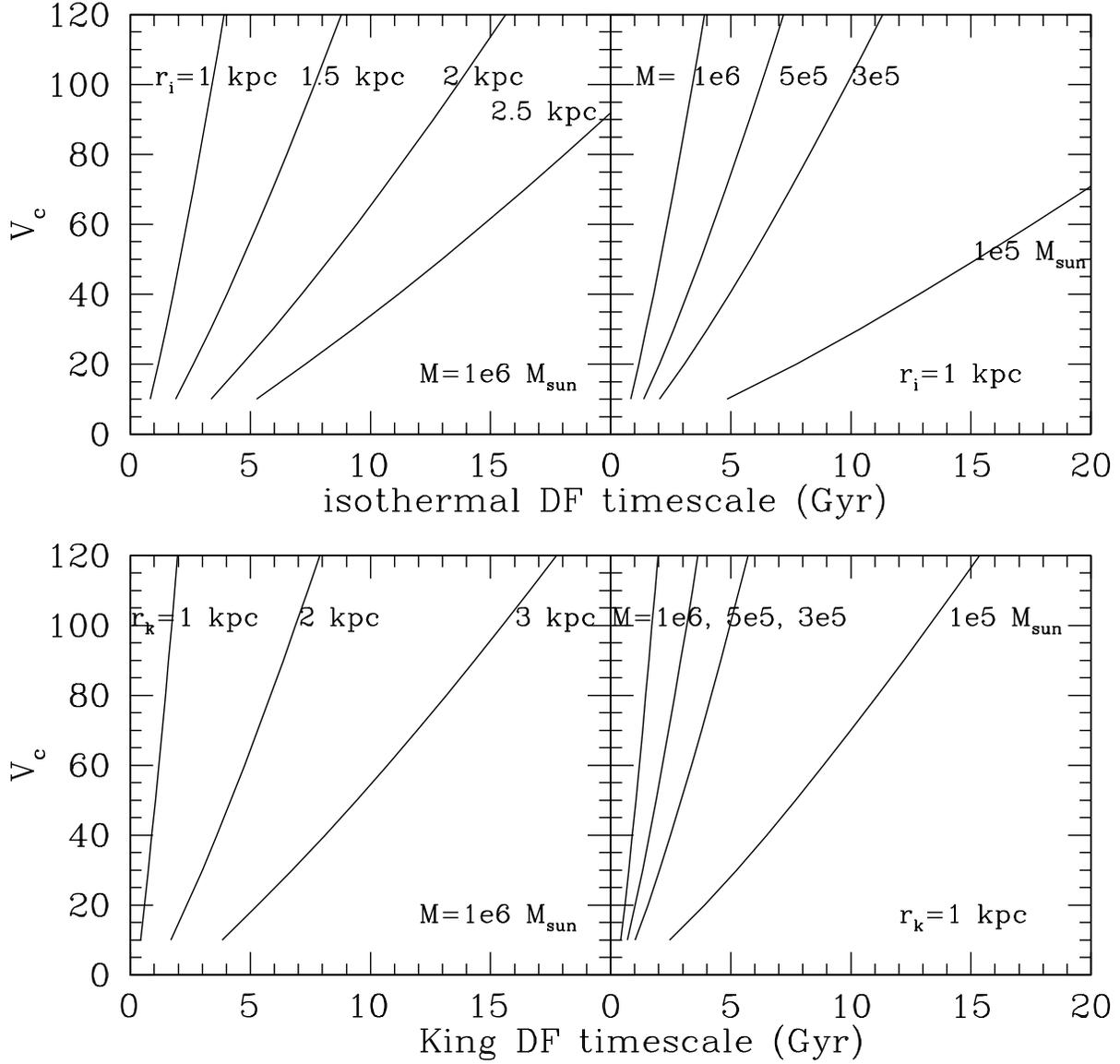}
\caption{The dynamical friction timescale for dwarf-sized isothermal
and King halos as a function of circular velocity.}
\end{figure}

\clearpage
\begin{figure}
\plotone{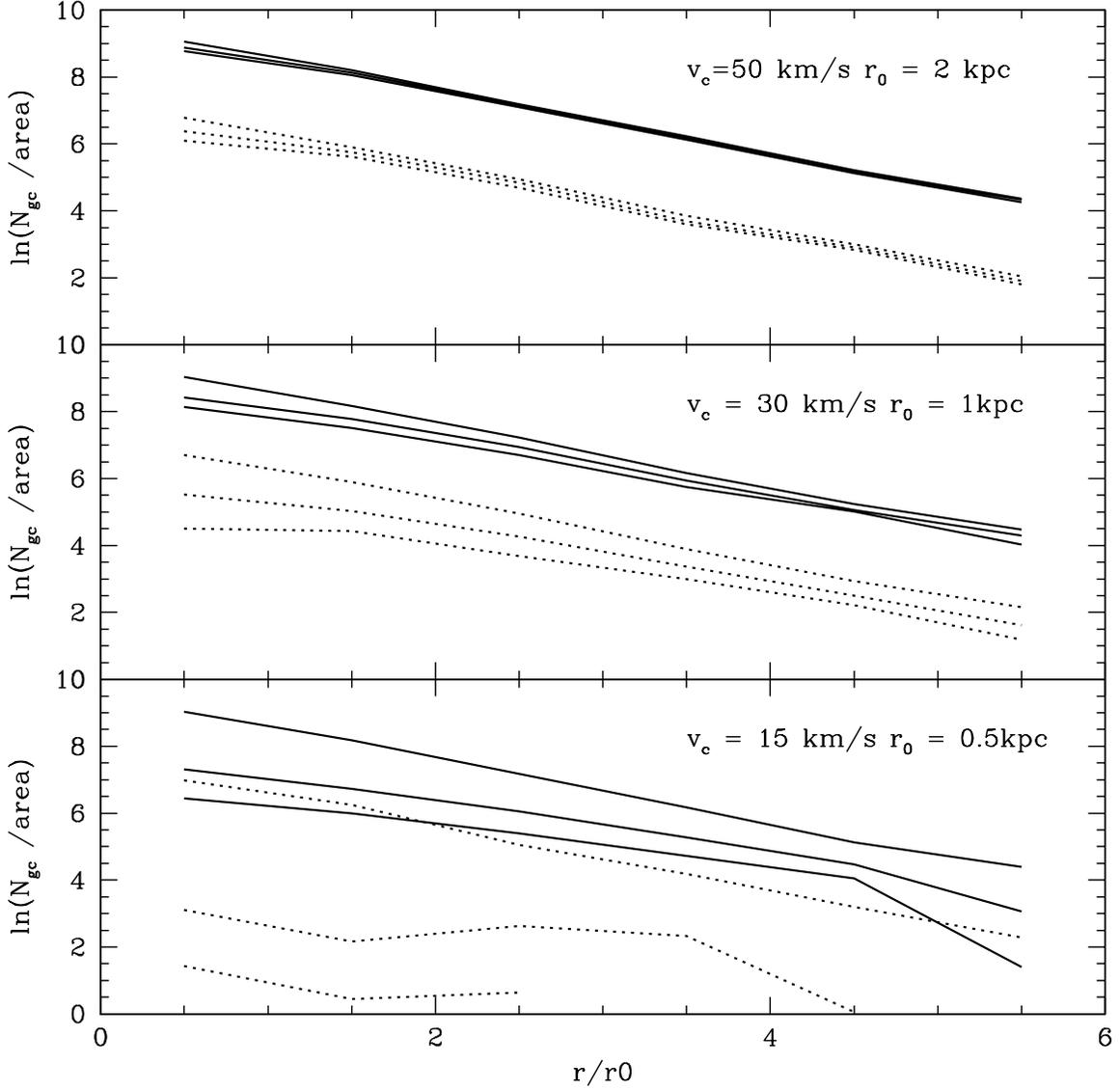}
\caption{The evolution of the entire (solid lines) and bright ($M_V < -8$)
(dashed lines) globular cluster radial profiles under the 
influence of dynamical friction in an isothermal halo 
after 0 (top line), 5 (middle), and 10 (bottom) Gyr.}
\end{figure}

\clearpage
\begin{figure}
\plotone{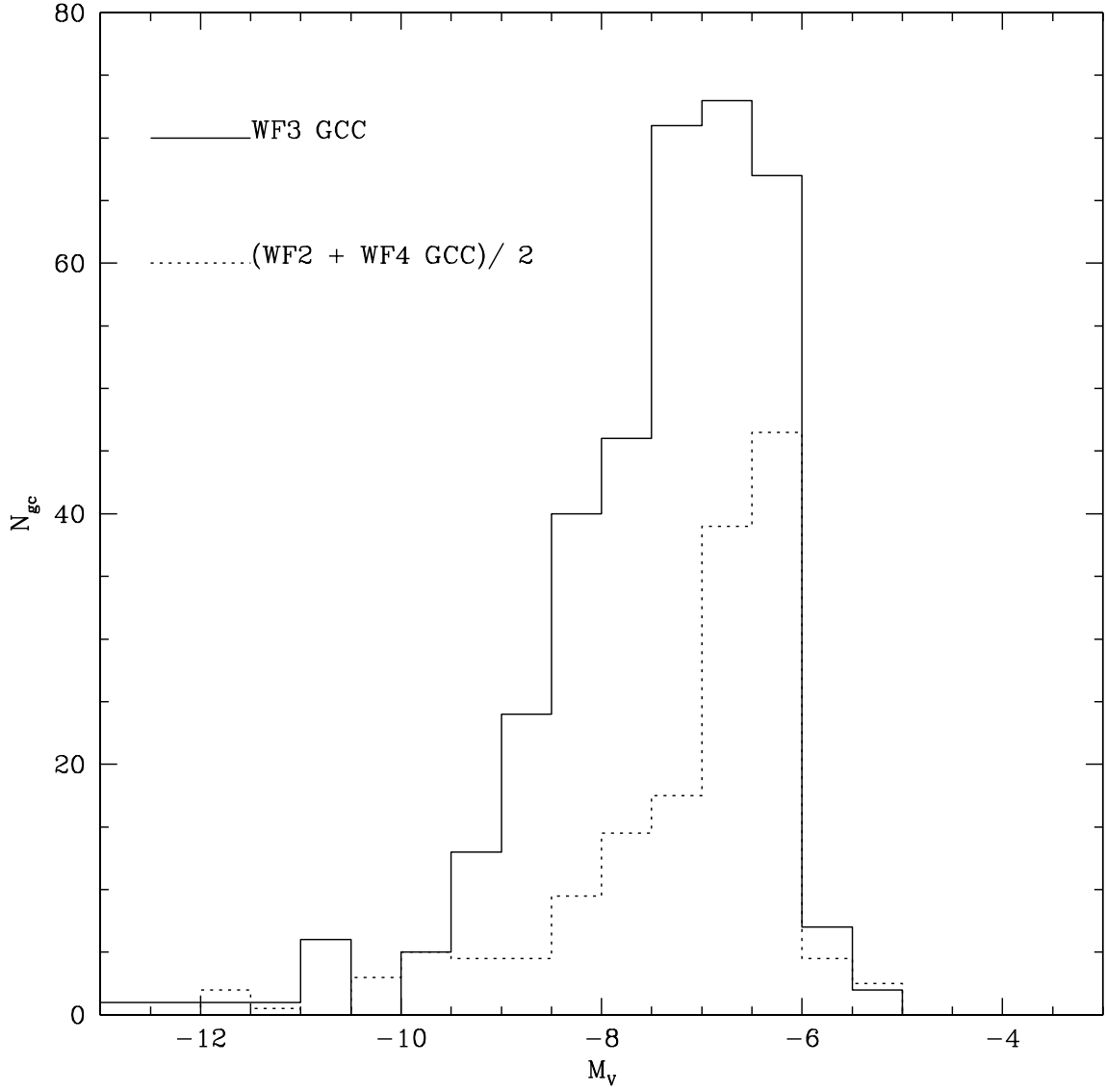}
\caption{The luminosity distribution of Fornax and Virgo dE
globular cluster candidates
(solid line) and background objects (dashed line) for which $0.5 <(V-I) <1.5$
and $FWHM < 2.5$ pixels.}
\end{figure}

\clearpage
\begin{figure}
\plotone{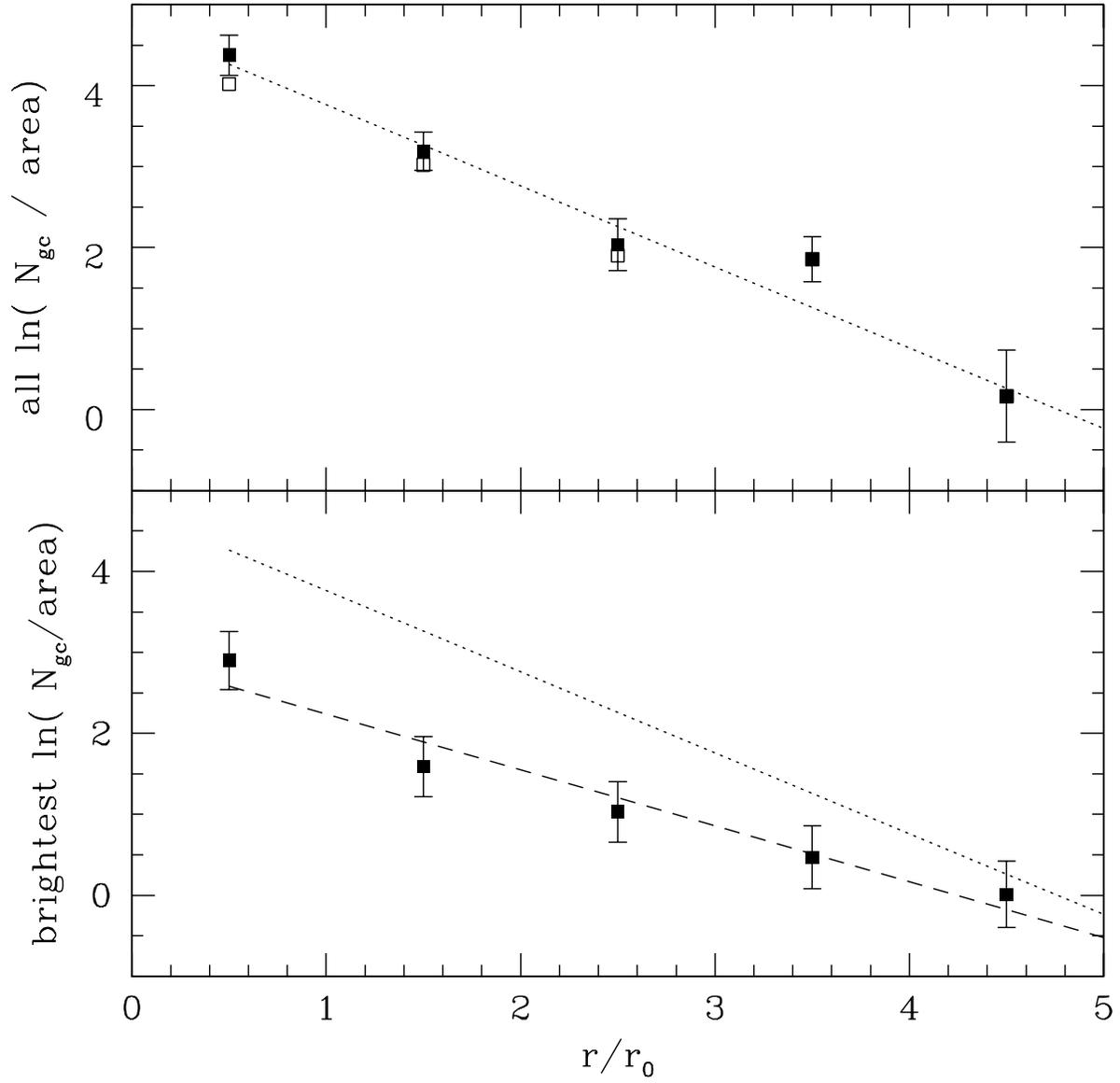}
\caption{The summed radial distribution of the globular cluster candidates, 
corrected for background contamination, scaled by scalelength of galaxy,
and excluding possible nuclei. The total globular
cluster surface density (upper panel) follows the exponential light profile of the 
dEs (dotted line), but the bright clusters ($M_V < -8.0$, lower panel)
 appear to be depleted and have
a shallower profile (dashed line). The open boxes in the top panel
show the summed profile prior to correction for incompleteness effects
in the inner regions. The error bars are $\sqrt{(N_{gc})^2 +
 (N_{background})^2 }/area$.}
\end{figure}

\clearpage
\begin{figure}
\plotone{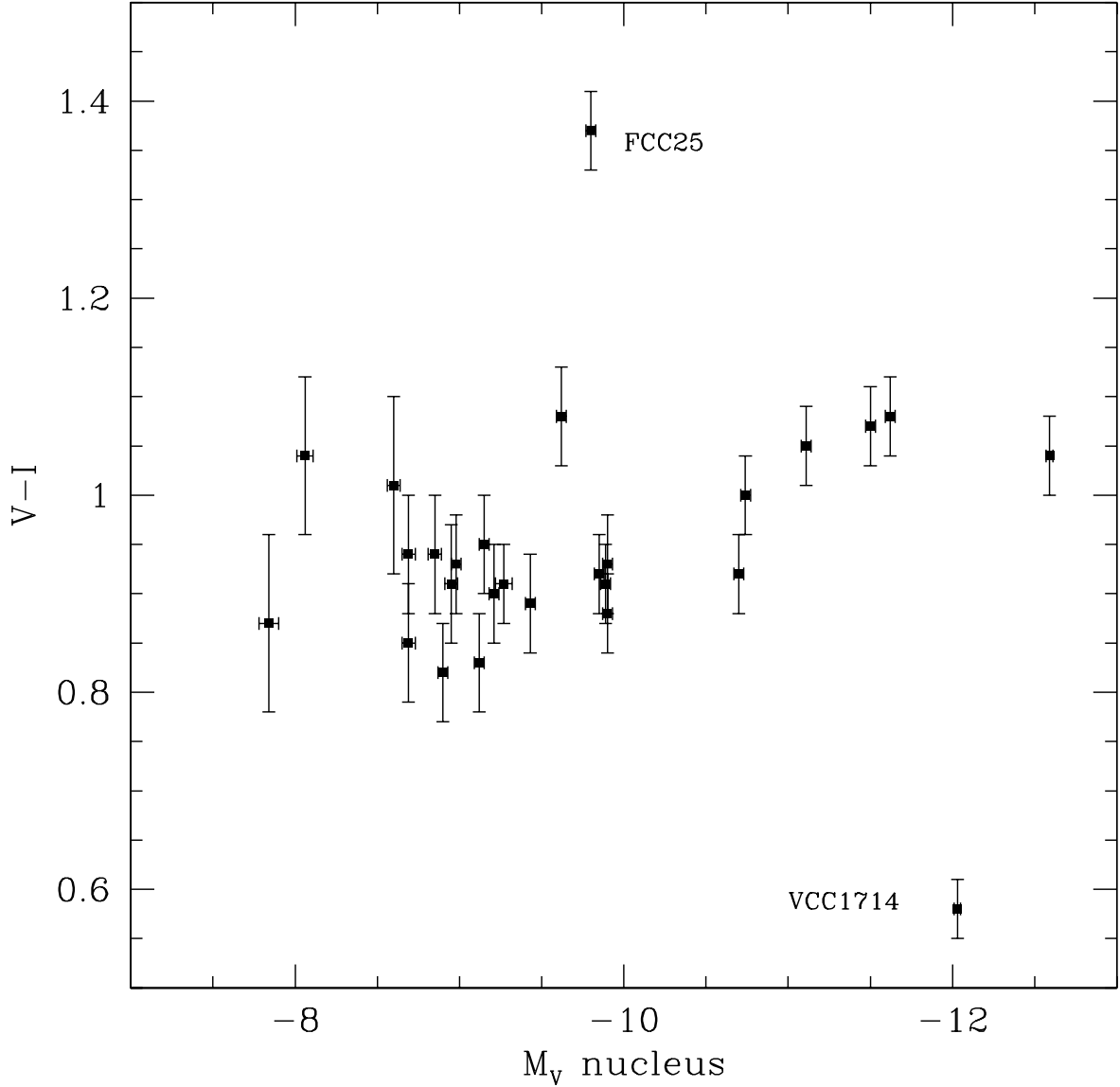}
\caption{$M_V$ vs. $V-I$ for the dE nuclei. Most nuclei have colors 
similar to old metal-poor Galactic globular clusters and have a slight
trend towards redder colors with increasing luminosity. The nucleus of FCC25
is $\sim$ 0.4 magnitude redder than the rest of the nuclei and could be
a background galaxy. The nucleus of VCC1714 is $\sim$ 0.3 magnitudes bluer
than the rest of the nuclei and is consistent with an age $<$ 1 Gyr.}
\end{figure}

\clearpage
\begin{figure}
\plotone{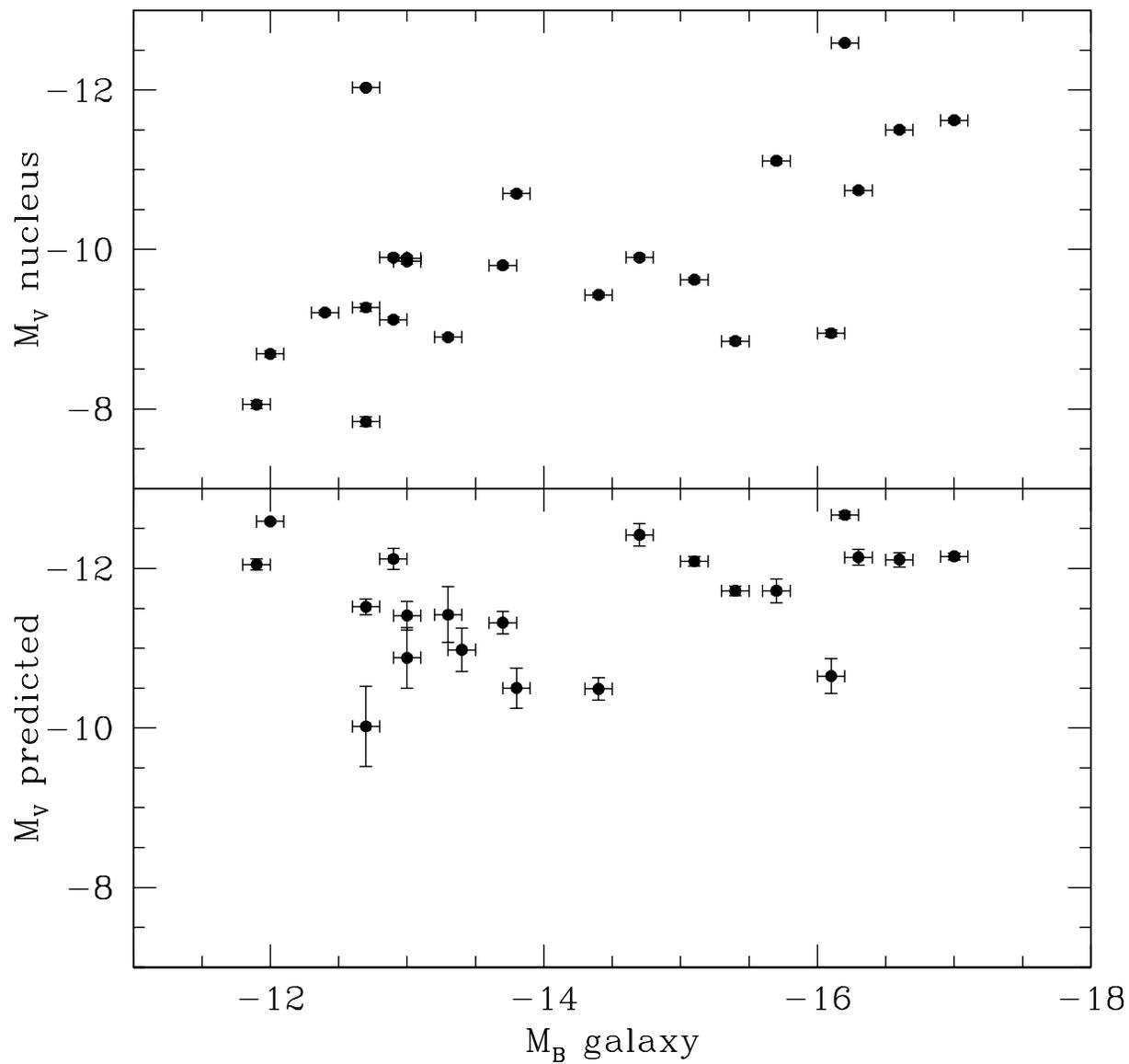}
\caption{$M_V$ of dE nuclei vs. $M_B$ of host galaxy.  Fainter galaxies
tend to have fainter nuclei (top).  Our simulation of dE nucleus formation
(bottom) predicts a much weaker trend of nuclear luminosity with host galaxy
luminosity.}
\end{figure}

\clearpage
\begin{figure}
\plotone{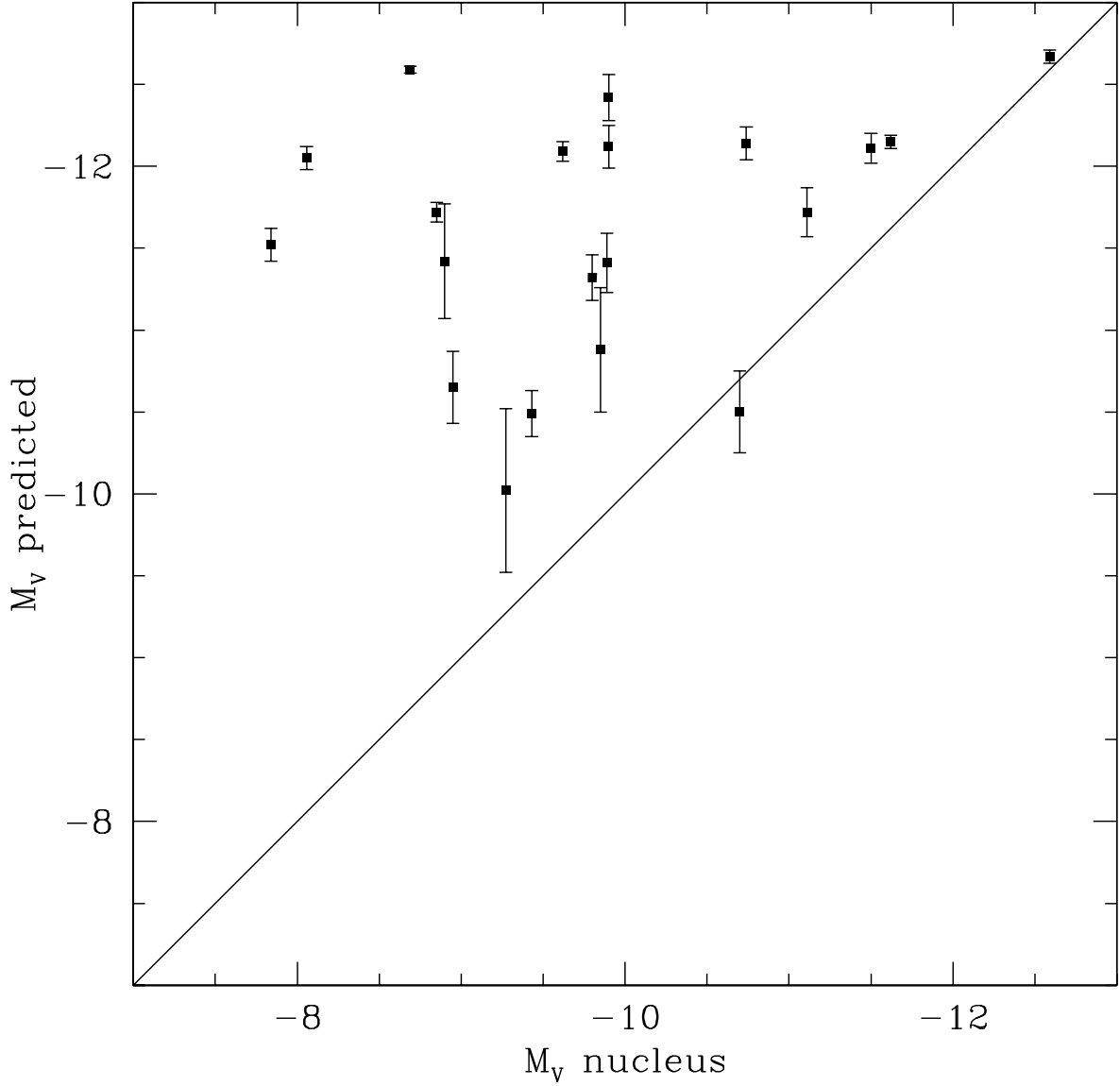}
\caption{The predicted nuclear $M_V$ from our Monte Carlo dynamical
friction simulation vs. the observed nuclear $M_V$ for each dE,N in our
sample.  We overestimate the nuclear $M_V$ for dE with $M_B > -14$ by several
magnitudes. The error bars are the standard deviation of the
predicted magnitude for 100,000 trials.  The standard deviation increases
for galaxies with smaller globular cluster systems because of the
random sampling of the globular cluster mass function.
} 
\end{figure}

\end{document}